\documentclass[aps,prd,twocolumn,secnumarabic,amssymb, amsmath,nobibnotes,showpacs]{revtex4-1}

\usepackage{amssymb,amsmath,amscd}
\usepackage{graphicx,calc,epsfig,pstricks,bbm}
\usepackage{tikz}
\usepackage{enumerate}
\usepackage{amsfonts}
\usepackage{amsmath}
\usepackage{array} 
\usepackage{dcolumn} 
\usepackage{longtable} 

\usepackage{mathtools}
\usepackage{color}
\usepackage{bm}
\usepackage{siunitx}
\usepackage{wrapfig}
\usepackage{amsthm}
\usepackage{slashed}
\usepackage{simplewick}
\usepackage{physics} 
\usepackage{mathtools}
\usepackage{tensor}
\usepackage{hhtensor}
\usepackage{fancyref}

\usepackage{pgfplots}

\usepackage{mathrsfs}
\DeclareMathAlphabet{\mathpzc}{OT1}{pzc}{m}{it}

\DeclarePairedDelimiterX\rbraket[2]{(}{)}{#1 \delimsize\vert #2}
\DeclarePairedDelimiterX\rdyad[2]{\lvert}{\rvert}{#1 \delimsize{)(} #2}
\newcommand{\un}{\underline}
\newcommand{\tn}{\tensor}

\newcommand{\bp}{\beta_{+}}
\newcommand{\bim}{\beta_{-}}

\newcommand{\cl}{\mathpzc}

\newcommand{\p}{\mathpzc{p}}

\newcommand{\oo}{$\ddot{\mathrm{o}}$}

\makeatletter
\newcommand{\leftlabel}[1]{&&
  \refstepcounter{equation}\ltx@label{#1}%
  \tagform@{\theequation}&&}
\makeatletter

\begin{document}
\title{Probabilistic interpretation of the wave function for the Bianchi I model}%

\author{Leonardo Agostini$^{1}$, Francesco Cianfrani$^{2}$, Giovanni Montani$^{1,3}$}%
\affiliation{$^{1}$
Physics Department, ``Sapienza'' University of Rome, P.le Aldo Moro 5, 00185 (Roma), Italy\\
$^2$ Institute
for Theoretical Physics, University of Wroc\l{}aw, Pl.\ Maksa Borna
9, Pl--50-204 Wroc\l{}aw, Poland\\
$^{3}$ ENEA, Fusion and Nuclear Safety Department, C.R. Frascati - Via E. Fermi, 45 (00044) Frascati (Roma), Italy.}
\date{\today}%

\begin{abstract}
We compare two different approaches for quantization of the Bianchi I model: a reduced phase space quantization, in which the isotropic Misner variables is taken as time, and the Vilenkin proposal, in which a semiclassical approximation is performed for the same variable. We outline the technical and interpretative issues of these two methods and we demonstrate that they provide equivalent results only if the dynamics is essentially dictated by the isotropic matter contribution.
\end{abstract}



\maketitle

\section{\label{sec:introduction}Introduction}

The Wheeler-Dewitt equation \cite{DeWitt:1967yk,DeWitt:1967ub,DeWitt:1967uc}, 
corresponding to the canonical quantization of the gravitational field in the
metric approach \cite{Kuchar1981}, is associated to a functional formalism and
it becomes a viable theory only in Minisuperspace, where the symmetry restriction 
reduces the dynamical problem to a finite number of degrees of freedom. The minisuperspace model
\cite{primordial cosmology,book} is a natural arena for the study of Quantum Cosmology, since, 
as soon as we consider homogeneous Universes (Bianchi models) \cite{Landau2},
the corresponding wave function is taken over a finite number of degrees of freedom,
typically the three cosmic scale factors (or some functions of them), determining the evolution of independent spatial
directions. 

However, the cosmological implementation of the Wheeler-DeWitt equation solves
the question concerning the viability of the canonical formalism, but it does
not address all the other shortcomings of 
this approach, like the absence of a dynamical Hilbert space as a general
feature and then the issue concerning the predictivity of 
the considered quantum theory \cite{primordial cosmology,Blyth:1975is}. Describing the Bianchi models via
Misner variables \cite{Misner:1969hg,Misner1971}, 
{\it i.e.} separating the isotropic volume component of the Universe from the
corresponding anisotropy degrees of freedom, 
we reach a very meaningful representation of the Bianchi model quantum
dynamics: a Klein-Gordon equation in the presence of a potential term in the coordinates. 
Indeed, the minisuper-metric is pseudo-Riemannian and the isotropic Misner
variable plays the role of a time-like coordinate, while the anisotropies are
space-like variables. What makes puzzling the physical characterization of a
generic Bianchi model (especially the type VIII and IX, which are the most general ones allowed by the
homogeneity constraint) is the impossibility to separate positive and negative
frequency solutions, which is a key-point of the original quantum physics prescription 
to get a positive defined probability density. 
This impossibility is due to the time dependence (dependence on the isotropic
Misner variable) of the Bianchi model potential (except for the Bianchi type I, where it vanishes), so that the
physical interpretation of the solution to the cosmological implementation of the
Wheeler-DeWitt equation is de facto forbidden. Actually, such difficulty in constructing a dynamical Hilbert space 
relies on a more general feature of canonical quantum gravity: the Wheeler-deWitt equation
is an equation of Klein-Gordon type, so that it has a conserved
current that does not lead to a positive definite inner product.
Two different approaches can be pursued to give a meaningful interpretation of the 
wave function of the Universe in terms of a probabilistic theory: 
\textit{i}) we can classically solve the Hamiltonian constraint
and then quantize the resulting Schr\"odinger equation \cite{Misner:1969ae,Benini:2006xu} (reduced phase space quantization \cite{Henneaux1992}, RPSQ); 
\textit{ii}) we assume, according to the Vilenkin proposal \cite{Vilenkin1989}, that the isotropic 
Misner variable approaches a quasi-classical limit, while the anisotropies remain pure quantum degrees of freedom. 
Even this latter approach leads to a Schr\"odinger equation, but, as we will see, it avoids the 
square root non-local Hamiltonian operator emerging from the former. 
What makes comparable the two approaches is actually the role of the time-like variable, 
say the internal time of the theory, played by the isotropic Misner variable
in the corresponding Schr\"odinger dynamics. 

However, despite its non-local character, the paradigm based on RPSQ is an exact procedure, 
requiring no WKB approximation on the Universe 
wave function, which on the contrary is a necessary step in the Vilenkin proposal. 

Thus, limiting our attention to the simple case of a Bianchi I cosmology in presence of matter (a time-dependent term), 
we here address a rigorous comparison of the two quantization methods, in order to 
determine under which restrictions the Vilenkin adiabatic WKB representation of 
the Universe volume dynamics becomes predictive. 

To this end, we carefully construct and then compare the wave functions of 
the Bianchi I model in the two cases, so clarifying which restrictions provide equal probabilities 
on the anisotropy variables.
The possibility for a comparison requires that the wave function evolution be
described via the same time parameter, {\it i.e.} 
in correspondence to the same functional form 
of the lapse functions. 
It is just this overlap of the time variables to impose the most restrictive
condition for the validity of the Vilenkin representation, {\it i.e.} the spectra of anisotropy
momenta must extend over small values with respect to the time-dependent matter term, which 
is necessary for the quasi-classical limit of the
isotropic Misner variable. 
This restriction is equivalent to say that the 
adiabatic WKB Vilenkin approach overlaps 
the quantum dynamics in RPSQ 
(also dubbed Arnowitt-Deser-Misner reduction of the Hamiltonian problem
\cite{Arnowitt:1962hi}), only if the anisotropy degrees of freedom are small and then the Universe is
mainly isotropic. 
By other words, we fix the impossibility to apply the Vilenkin quantum
evolution of the Universe anisotropies near the cosmological singularity, 
where their values and momenta are arbitrarily large.

Nonetheless, once this strong restriction is 
fulfilled, we demonstrate how the probabilities (taken on the same domain) for the Universe
anisotropies coincide in the two approaches, suggesting that 
for a quasi-isotropic Universe a robust 
probabilistic interpretation exists. 
We conclude observing how, in the absence of a clear mechanism for frequency separation, 
the procedure of solving the Hamiltonian constraint and then canonically quantizing it remains the only
general attempt to the minisuperspace quantum dynamics, although it suffers the
aforementioned non-trivial question concerning locality. 

\section{Action for Bianchi models}

Let us consider \textit{homogeneous} but not necessarily \textit{isotropic} spacetimes. Following the definition given in \cite{Wald1984}, spatially homogenous spacetimes are those that can be foliated with a family of three-dimensional hypersurfaces $\Sigma_t$ and there exists an isometry of the three-metric $h_{ij}$ that connects any two points of $\Sigma_t$. For such spacetimes, the three-metric can be written as \cite{Landau2,primordial cosmology}

\begin{equation}
\label{eq:three_metric_homogeneous}
h_{ij}(x^k, t) = h_{ab}(t) e^{(a)}_i(x^k) e^{(b)}_j(x^k),
\end{equation}
where the vectors $e^{(a)}_i$ constitutes the so-called \textit{frame} and they do not depend on time. Homogeneous spaces have been  classified by Bianchi in \cite{Bianchi1898}. One can write $h_{ab} = (e^{2 \alpha} e^{2 \vb*{\beta}})_{ab}$ where $\beta_{ab}$ is a $3 \times 3$ traceless, symmetric matrix and $\alpha$ is called the isotropic variable, both depending on $t$ only. 

It is generically assumed that the supermomentum constraint identically vanishes and one can safely fix $N^i=0$ (see \cite{Moniz2010} for more details).
Furthermore, $\beta$ can be diagonalized and the eigenvalues can be parametrized as follows

\begin{equation}
\label{eq:beta_matrix_diagonal}
\beta_{ab} = diag( \bp + \sqrt{3} \bim, \bp - \sqrt{3} \bim, - 2 \bp).
\end{equation}
Making a canonical transformation to $\alpha, \bp, \bim$, their conjugate momenta $\cl{p}_{\alpha}, \cl{p}_+, \cl{p}_-$ are given by 
\begin{align}
&\p_{\alpha}=- \frac{3 c^2 \kappa e^{3 \alpha}}{4 \pi G N}\,\dot{\alpha} \\
&\p_+ = \frac{3 c^2 \kappa e^{3 \alpha}}{4 \pi G N}\,\dot{\beta}_+ \\
&\p_-= \frac{3 c^2 \kappa e^{3 \alpha}}{4 \pi G N}\,\dot{\beta}_- \,.
\end{align}

and the Einstein-Hilbert action becomes

\begin{equation}
\label{eq:action_alpha_beta}
\begin{aligned}
S_{EH}=& \int \dd t \Bigg\{ \cl{p}_{\alpha} \dot{\alpha} + \cl{p}_+ \dot{\beta}_+ + \cl{p}_- \dot{\beta}_- - \frac{2 \pi G N e^{-3 \alpha}}{3 c^2 \kappa} \\
&\qty[ - \cl{p}_{\alpha}^2 + \cl{p}^2_+ + \cl{p}^2_- - \frac{3 c^2 \kappa^2 e^{6 \alpha}}{32 \pi^2 G^2} \tn[^{(3)}]{R}{}] \Bigg\}
\end{aligned}
\end{equation}
where $\kappa = \int \dd^3 x \abs{\det(e^{(a)}_i(x^k))}$. Let us add as matter terms some perfect fluids with pressure $P$ and energy density $\rho$ having the equation of states $P=w\,\rho$. The total action reads as

\begin{equation}
\label{eq:action_alpha_beta_+_matter}
\begin{aligned}
S=& \int \dd t \Bigg\{ \cl{p}_{\alpha} \dot{\alpha} + \cl{p}_+ \dot{\beta}_+ + \cl{p}_- \dot{\beta}_- - \frac{2 \pi G N e^{-3 \alpha}}{3 c^2 \kappa} \\
&\qty[ - \cl{p}_{\alpha}^2 + \cl{p}^2_+ + \cl{p}^2_- - \frac{3 c^2 \kappa^2 e^{6 \alpha}}{32 \pi^2 G^2} \tn[^{(3)}]{R}{} + \mu^2(\alpha)] \Bigg\}
\end{aligned}
\end{equation}
where the matter contribution is encoded in the term $\mu^2$, reading, according to \cite{Ryan1975}, 

\begin{equation}
\label{eq:matter_term} 
\begin{aligned}
\mu^2(\alpha)= \sum_w \mu^2_w e^{3(1-w)\alpha}, 
\end{aligned}
\end{equation}
and the sum extends over all the fluid components, characterized by different values of $w$, while $\mu^2_w$ are constants. 
 
Let us now consider the easiest case of the Bianchi I model, for which $\tn[^{(3)}]{R}{} = 0$ and we obtain:

\begin{equation}
\label{eq:action_for_bianchi_I}
\begin{aligned}
S=& \int \dd t \Bigg\{ \cl{p}_{\alpha} \dot{\alpha} + \cl{p}_+ \dot{\beta}_+ + \cl{p}_- \dot{\beta}_- - \frac{2 \pi G N e^{-3 \alpha}}{3 c^2 \kappa} \\
&\qty[ - \cl{p}_{\alpha}^2 + \cl{p}^2_+ + \cl{p}^2_- + \mu^2(\alpha)] \Bigg\}.
\end{aligned}
\end{equation}
Notice that Eq. \eqref{eq:action_for_bianchi_I} resembles the action of a relativistic particle moving in a $(\alpha, \bp, \bim)$ space with a variable mass. Moreover, it is worth saying that $\kappa$ can be made finite even if we are considering a flat space, closing it with a torus topology or, due to homogeneity, just considering a finite portion of space. From Eq. \eqref{eq:action_for_bianchi_I} we can find the classical equations of motion:

\begin{subequations}
\label{eq:classical_equation_of_motion}
\begin{align}
&- \dv{\mu^2(\alpha)}{\alpha} - \frac{3 c^2 \kappa}{2 \pi G N} e^{3 \alpha} \dot{\p}_{\alpha}=0; \label{eq:motion_for_p_alpha}\\
&\dot{\p}_+ = 0, \quad \dot{\p}_- = 0; \label{eq:motion_for_p_+_-}\\
&\dot{\alpha} = - \frac{4 \pi G N e^{-3 \alpha}}{3 c^2 \kappa} \p_{\alpha}; \label{eq:motion_for_alpha}\\
&\dot{\beta}_+ = \frac{4 \pi G N e^{-3 \alpha}}{3 c^2 \kappa} \p_+; \label{eq:motion_for_bp}\\
&\dot{\beta}_- = \frac{4 \pi G N e^{-3\alpha}}{3 c^2 \kappa} \p_-; \label{eq:motion_for_bim}\\
&\mathcal{H} = (- \p^2_{\alpha} + \p_+^2 + \p_-^2) + \mu^2(\alpha)=0. \label{eq:first_class_constraint}
\end{align}
\end{subequations}
It is worth noticing how momenta $\p_{+}$ and $\p_{-}$ are constants of motion.


\section{RPSQ of Bianchi I}

The idea of RPSQ is to reduce the phase-space to the only physical d.o.f. and to choose a time variable \textit{before} quantising the theory. It is worth noticing that we are dealing with $\alpha, \bp$ and $\bim$ and then one variable is to some extent redundant. This is due to the fact that we still have the gauge freedom of the lapse function form, which can be fixed in order to set $\alpha$ as a time-like variable. Similarly to \cite{Misner1971}, we here make the gauge choice $\alpha = t/T := \cl{t}$, providing that $\dot{\alpha} > 0$ in order to select the expanding branch of the Universe. This implies $\p_{\alpha} < 0$ since $N$ is positive defined. The constant $T$ can be defined in terms of fundamental constants (in which case it can be chosen proportional to the Planck length), or it can be derived from the specific data of the considered problem (as soon as a proper matter or energy scale is given from the matter fields configuration). The lapse function is then fixed: from \eqref{eq:motion_for_alpha} we get

\begin{equation}
\label{eq:lapse_function_adm}
N_{RPSQ} (\cl{t}) = - \frac{3 c^2 \kappa}{4 \pi G T}\,\frac{e^{3 \cl{t}}}{\cl{p}_{\alpha}}%
\end{equation}

In order to reach a fully reduced phase-space, we classically solve the first class constraint \eqref{eq:first_class_constraint} obtaining

\begin{equation}
\label{eq:solving_firt_class_constraint}
\cl{p}_{\alpha} = -\sqrt{\cl{p}_+^2 + \cl{p}_-^2 + \mu^2(\alpha)}.
\end{equation}
Then we substitute \eqref{eq:solving_firt_class_constraint} and \eqref{eq:lapse_function_adm} into \eqref{eq:action_for_bianchi_I} obtaining:

\begin{equation}
\label{eq:rpsq_action}
S_{RPS} = \int \dd \cl{t} \qty[ \cl{p}_+ \dv{\bp}{\cl{t}} + \cl{p}_- \dv{\bim}{\cl{t}} - \sqrt{\cl{p}_+^2 + \cl{p}_-^2 + \mu^2(\cl{t})} ]
\end{equation}
We have reached a form for the action in which we have only the two physical d.o.f. and a nonlocal Hamiltonian, corresponding to a square root operator. We can therefore proceed with the canonical quantisation by introducing the wave function $\Phi(\bp,\bim,\cl{t})=\Phi(\un{\beta},\cl{t})$ and promoting momenta to derivative operators as follows
 
 \begin{equation}
 \label{eq:canonical_quantisation_rps}
 \cl{p}_+ \rightarrow - i \hbar \pdv{\bp}, \quad \cl{p}_- \rightarrow - i \hbar \pdv{\bim}\,,
 \end{equation}
and then we get the following Schr{\oo}dinger equation:

\begin{equation}
\label{eq:schrodinger_equation_rps}
i \hbar \pdv{\cl{t}} \Phi(\un{\beta},\cl{t}) = \sqrt{- \hbar^2 \qty(\pdv[2]{\bp} + \pdv[2]{\bim}) + \mu^2(\cl{t})}\, \Phi(\un{\beta},\cl{t})\,.
\end{equation}
The equation above resembles the spinless Salpeter equation \cite{Lammerzahl1993,Schweber1961}, except for the presence of the time-dependent mass term $\mu^2(\cl{t})$. It is worth noting that for $w\leq 1$ the function $\mu^2(\cl{t})$ is bounded close to the singularity, corresponding to $\alpha\rightarrow -\infty$, and we denote by $\mu^2_1$ the corresponding value, {\it i.e.}
$\lim_{\cl{t} \to - \infty} \mu^2(t) = \mu_1^2$. Hence, close to the singularity the Schr{\oo}dinger equation coincides with the Salpeter equation with mass term $\mu_1^2$. 
Generically, the operator under square root on the right-hand side of \eqref{eq:schrodinger_equation_rps} is \textit{strongly elliptic} (see \cite{Lammerzahl1993}), therefore the square root operator is a \textit{pseudo-differential operator}, which can be conveniently analyzed in the base of plane waves 
$\phi_{\un{p}}(\un{\beta}) = A(\un{p}) e^{\frac{i}{\hbar} \un{p} \cdot \un{\beta}}$ finding

\begin{equation}
\label{eq:plane_wave_solution}
\sqrt{- \hbar^2 \Delta_{\pm} + \mu^2(\cl{t})} \phi_{\un{p}}(\un{\beta}) = \sqrt{\un{p}^2 + \mu^2(\cl{t})} \phi_{\un{p}}(\un{\beta}),
\end{equation}
where $\un{p}^2 = p_+^2 + p_-^2$. Hence, the evolution of plane waves is described by

\begin{equation}
\label{eq:evolution_of_phi_transf}
\tilde{\Phi}(\un{p},\cl{t})= e^{- \frac{i}{\hbar} \int_{\cl{t}_0}^{\cl{t}} \dd \cl{t}' \sqrt{\abs{\un{p}}^2 + \mu^2(\cl{t})}} \tilde{\Phi}(\un{p},\cl{t}_0)\,,
\end{equation}
and via inverse Fourier transform a generic solution can be formally written as

\begin{equation}
\label{eq:evolution_of_phi}
\Phi(\un{\beta},\cl{t})= e^{- \frac{i}{\hbar} \int_{\cl{t}_0}^{\cl{t}} \dd \cl{t}' \sqrt{-\hbar^2 \Delta_{\pm}^2 + \mu^2(\cl{t})}} \Phi(\un{\beta},\cl{t}_0).
\end{equation}
The main issue of this formulation is the definition of a suitable inner product. A Klein-Gordon-like inner product is an attractive possibility, given the results in \cite{Schweber1961}, but here it would be intrinsically time-dependent, because of the presence of the time-dependent mass term $\mu^2(t)$. Hence, we propose to specify the inner product near the singularity ($\cl{t} \rightarrow - \infty$), where $\mu^2\rightarrow \mu_1^2$, and to define it generically in a time-independent way, according to \cite{Mostafazadeh:2002xa}, so finding

\begin{equation}
\label{eq:inner_product_sqrt_kg_infty}
\begin{aligned}
\rbraket{\psi}{\phi}(\cl{t}) =& \int \dd^2 \beta \bigg[ \psi^*(\un{\beta},\cl{t}) \,\left(\sqrt{-\hbar^2 \Delta_{\pm} + \mu^2_1}\, \phi(\un{\beta},\cl{t})\right) +\\
&+ \left(\sqrt{-\hbar^2 \Delta_{\pm} + \mu^2_1}\, \psi^*(\un{\beta},\cl{t})\right) \,\phi(\un{\beta},\cl{t})\bigg].
\end{aligned}
\end{equation}
It can be shown that the evolution operator in \eqref{eq:evolution_of_phi} is \textit{unitary} with respect to such an inner product. This also implies that, if we normalize the wave function in the limit $\cl{t} \rightarrow -\infty$, the normalization is preserved for any $\cl{t}$.


\section{Semiclassical approximation}

We now adopt the Dirac quantisation method in order to obtain a quantum theory describing the Bianchi I cosmological model, {\it i.e.}, instead of classically solving the super-Hamiltonian constraint, we quantise it requiring that it annihilates physical state $\ket{\Psi}$

\begin{equation}
\label{eq:constraint_operator_on_physical_states}
\vu*{\mathcal{H}} \ket{\Psi} = 0.
\end{equation}
The canonical quantisation procedure can be implemented similarly to the RPSQ case \eqref{eq:canonical_quantisation_rps}, but in addition we also have to implement $\alpha$ as a multiplicative operator and $\p_{\alpha} \rightarrow -i \hbar \pdv{\alpha}$. Hence, the condition \eqref{eq:constraint_operator_on_physical_states} becomes

\begin{equation}
\label{eq:equation_for_psi}
\qty[ \hbar^2 \qty( - \pdv[2]{\alpha} + \pdv[2]{\bp} + \pdv[2]{\bim}) + \mu^2(\alpha)] \Psi(\un{\beta},\alpha)=0.
\end{equation}

Since $\vu*{\mathcal{H}}$ is the generator of gauge transformations, \eqref{eq:constraint_operator_on_physical_states} ensures that on a quantum level physical states do not change under gauge transformations, {\it i.e.}, the wavefunction does not depend on the label time $t$. 
Therefore, the time is hidden between the internal variables $\alpha, \bp, \bim$. Focusing on the probabilistic interpretation for the wave function, De Witt \cite{DeWitt:1967yk} first suggested to define a scalar product from a conserved current $j^{\mu}$, analogously to the Klein-Gordon case. Although a Klein-Gordon-like current can be defined, the corresponding scalar product is not positive defined and no separation of frequencies can be consistently implemented (this is due to a divergence occurring in superspace when the metric determinant vanishes, see for instance \cite{book}). 

A different approach has been proposed by Vilenkin \cite{Vilenkin1989}, who suggested that a proper probabilistic interpretation can be achieved only after the emergence of time via a semiclassical approximation on the wave function $\Psi$. Making the ansatz

 \begin{equation} 
 \label{eq:ansatz_semiclassical}
 \Psi(\un{\beta},\alpha) = A(\alpha) e^{\frac{i}{\hbar} S(\alpha)} \chi(\un{\beta}, \alpha),
 \end{equation}
it can be shown that a positive defined probability exists for $\chi$ if the congruence of classical trajectories ({\it i.e.} the gradient curves of $S_0$) crosses once and only once the hypersurfaces of constant timelike variable. In particular, $\alpha$ is the time-like variable and the contour lines of $S(\alpha)$ are equal-time hypersurfaces. 
Inserting the wave function \eqref{eq:ansatz_semiclassical} into \eqref{eq:equation_for_psi}, a solution is given for
 
 \begin{subequations}
 \label{eq:eqs_vilenkin}
 \begin{align}
 &-\qty( \dv{S(\alpha)}{\alpha})^2 + \mu^2(\alpha) = 0; \label{eq:eq_for_s_alpha}\\
 &\dv{\alpha} \qty[ A^2 (\alpha) \qty(- \dv{S(\alpha)}{\alpha})]=0; \label{eq:classical_continuity_eq} \\
 &- i \hbar \dv{S(\alpha)}{\alpha} \pdv{\chi(\un{\beta},\alpha)}{\alpha} = - \frac{\hbar^2}{2} \qty( \pdv[2]{\bp} + \pdv[2]{\bim}) \chi(\un{\beta},\alpha). \label{eq:approximated_eq_chi}
 \end{align}
 \end{subequations}

In particular, from \eqref{eq:eq_for_s_alpha} and \eqref{eq:classical_continuity_eq} one gets

\begin{subequations}
\label{eq:solution}
\begin{align}
&S(\alpha) =  \pm \int_{\alpha_0}^{\alpha} \dd \alpha' \sqrt{\mu^2(\alpha')} \\
&A(\alpha) = \frac{\mathcal{K}^2}{\sqrt[4]{\mu^2(\alpha)}}. 
\end{align}
\end{subequations}
where $\mathcal{K}^2$ is an arbitrary integration constant. By introducing the WKB time $t_{WKB}$, for which $\dv{S(\alpha)}{\alpha} \pdv{\alpha} = \pdv{t_{WKB}}$ \cite{Kiefer1987}, \eqref{eq:approximated_eq_chi} becomes the Schr{\oo}dinger equation describing the motion of a free particle in the $(\bp,\bim)$ plane. Therefore, $\chi$ can be interpreted as the wave function describing a free particle in superspace, in which $\alpha$ is a parametric time-like variable.   

By solving the eigenvalue problem for \eqref{eq:approximated_eq_chi}, redefining $\chi(\un{\beta},\alpha)$ in order to set $\mathcal{K}^2=1$ and normalizing $\chi(\un{\beta},\alpha)$ with respect to the natural scalar product for a free particle, {\it i.e.}

\begin{equation}
\label{eq:inner_product_vlienkin}
\braket{\chi(\alpha)}{\chi(\alpha)} = \int \dd^2 \beta | \chi(\un{\beta},\alpha) |^2=1,
\end{equation}
we get the following result 

\begin{equation}
\label{eq:final_result_vilenkin}
\begin{aligned}
\Psi(\alpha,\un{\beta}) =& \frac{e^{- \frac{i}{\hbar} \int^{\alpha}_{\alpha_0} \dd \alpha' \sqrt{\mu^2(\alpha)}}}{\sqrt[4]{\mu^2(\alpha)}}\\
& \int_{\mathbb{R}^2} \frac{\dd^2 p}{2 \pi \hbar} e^{- \frac{i}{2\hbar} (p^2_+ + p^2_-) \int^{\alpha}_{\alpha_0} \dd \alpha' \frac{1}{\sqrt{\mu^2(\alpha)} }} \cdot \\ 
&\cdot e^{\frac{i}{\hbar} \un{p} \cdot \un{\beta}} \tilde{\chi}(\alpha_0, \un{p}).\\
\end{aligned}
\end{equation}
where the function $\tilde{\chi}(\alpha_0, \un{p})$ determines initial conditions. It is worth saying that two hypotheses lie behind Vilenkin's semiclassical approximation: a WKB approximation and a Born-Oppenheimer approximation, in which $\alpha$ is the slow variable whereas $\bp$ and $\bim$ are the fast ones. The validity of these approximations implies that 

\begin{enumerate}
\item we admit a decomposition of the wave function as $\Psi(\alpha,\un{\beta})=e^{\frac{i}{\hbar}\sum_{n=0} (\hbar)^n\,S_{(n)}}$
and the following conditions hold 
\begin{subequations}
\label{eq_bianchiI:semiclassical_conditions_final}
\begin{align}
&\bullet \ \abs{\frac{1}{\mu^{3}(\alpha)} \dv{\mu^2(\alpha)}{\alpha}} \ll \frac{4}{\hbar}; \label{eq_bianchiI:first_semiclassical_condition_final}\\
&\bullet \ \hbar \abs{S_2(\alpha)} \ll |S_1(\alpha)| \quad \mbox{and} \quad \hbar \abs{S_2(\alpha)} \ll 1. \label{eq_bianchiI:second_semiclassical_condition_final}
\end{align}
\end{subequations}
\item the integral over $p$s in \eqref{eq:final_result_vilenkin} extends over those values for which 
\begin{subequations}
\label{eq_bianchiI:Born-Oppenheimer_condition_final}
\begin{align}
&\bullet \ (p_+^2+p_-^2) \ne 0; \label{eq_bianchiI:Born-Oppenheimer_non_zero}\\
&\bullet \ (p_+^2 + p_-^2) \ll \mu^2 (\alpha). \label{eq_bianchiI:Born-Oppenheimer_lesser_than_h_m}
\end{align}
\end{subequations}
\end{enumerate}

From \eqref{eq:matter_term} it can be shown that there always exists $\alpha_S$ such that for $\alpha>\alpha_S$ the WKB approximation is satisfied, whereas the BO approximation implies that we need a wave packet $\tilde{\chi}(\alpha_0, \un{p})$ sufficiently peaked near an \textit{ad hoc} value $\bar{\un{p}}=(\bar{\un{p}}_+,\bar{\un{p}}_-)$ for which \eqref{eq_bianchiI:Born-Oppenheimer_lesser_than_h_m} holds. For simplicity, we choose a Gaussian wave packet. It is worth noting that since the Hamiltonian operator is a function of momentum operators only, no wave function spread occurs in the $p$-representation. This ensures that condition \eqref{eq_bianchiI:Born-Oppenheimer_lesser_than_h_m} is satisfied for every $\alpha$ if it is satisfied at a certain $\alpha_0$.  Thus we get the following final expression for the wave function in semiclassical approximation:

\begin{equation}
\label{eqn_bianchiI:final_wfv_fin}
\begin{aligned}
\Psi(\alpha,\un{\beta}) =& \frac{e^{- \frac{i}{\hbar} \int^{\alpha} \dd \alpha' \sqrt{\mu^2(\alpha)}}}{\sqrt[4]{\mu^2(\alpha)}} \\
& \int_{\mathbb{R}^2} \frac{\dd^2 p}{2 \pi \hbar} \frac{e^{\frac{i}{2\hbar} (p^2_+ + p^2_-) \int^{\alpha} \dd \alpha' \frac{1}{\sqrt{\mu^2(\alpha)} }}}{\sqrt{\pi \sigma_+ \sigma_-}} \cdot \\ 
&\cdot e^{-\frac{i}{\hbar} \un{p} \cdot (\un{\cl{b}}-\un{\beta})} e^{- \frac{(p_+ - \bar{\un{p}}_+)^2}{2 \sigma^2_+}} e^{- \frac{(p_- - \bar{\un{p}}_-)^2}{2 \sigma^2_-}}\,,\\
\end{aligned}
\end{equation}
where $\un{\cl{b}}=(\un{\cl{b}}_+,\un{\cl{b}}_-)$ denotes the point in the $(\bp,\bim)$ plane around which the wave function is initially peaked.
For the following analysis, we write here the explicit expression of the lapse function $N_S$, that comes out from the fact that $p_{\alpha} = \dv{S(\alpha)}{\alpha}$ and from equation \eqref{eq:motion_for_alpha}, namely

\begin{equation}
\label{eq:lapse_semiclassical}
N_S(\alpha) = \frac{3 c \kappa \dot{\alpha} e^{3 \alpha}}{4 \pi G} \frac{1}{\sqrt{\mu^2(\alpha)}}.
\end{equation}


\section{Comparison}

In order to compare the two formulations, we first need to identify the same time variable. It can be done requesting that $\alpha=t/T = \cl{t}$ holds also for the semiclassical approximation and asking for the same label time $t$ in the two formulations, \textit{i.e.}, we need the two lapse function \eqref{eq:lapse_semiclassical} and \eqref{eq:lapse_function_adm} to be the same
%

\begin{equation}
\label{eq_bI:comparison_lapse_w_1_approx}
\frac{3 c \kappa}{4 \pi G T} \frac{e^{3 \cl{t}}}{\sqrt{\mathpzc{p}^2_+ + \mathpzc{p}^2_- + \mu^2(\cl{t})}} = \frac{3 c \kappa}{4 \pi G T} \frac{e^{3 \cl{t}}}{\sqrt{\mu^2(\cl{t})}},
\end{equation}
This equation is essentially valid only if $\cl{p}_+^2 + \cl{p}_-^2 \ll \mathcal \mu^2(t)$, but this request coincides with \eqref{eq_bianchiI:Born-Oppenheimer_lesser_than_h_m}. 

We now observe that once we promote $\bp$ and $\bim$ to quantum operators, the lapse function $N_{RPSQ}$ \eqref{eq:lapse_function_adm} becomes itself an operator acting on the wave function \cite{Arnowitt:1962hi,Misner:1969ae}. Thus, in order to perform the comparison above, we have to replace it by its expectation value on the wave function, essentially corresponding to its expression calculated on the classical momenta $\bar{\un{p}}$ (this requires, as addressed below, that the wave function is sufficiently peaked around the classical trajectories). However, it is worth noting that for the case of Bianchi I $\bar{\un{p}}$ are classical constants of motion and therefore we can think of them as fixed numbers in the expression of the RPSQ lapse function. Around these fixed values, we can then peak the Universe wave function, in order to perform the comparison \eqref{eq_bI:comparison_lapse_w_1_approx}.

Our aim is to check whether the functional form of the wave functions obtained in the two considered cases, or their associated probabilities, coincide. In doing this matching, we need to take a range of $\cl{t}$ such that the semiclassical approximation is satisfied. Calling $\cl{t}_S$ such time, we also impose that $\cl{t}_0 \ge \cl{t}_S$ and we consider initial conditions providing the same probability to find the Universe in a point $(\un{\cl{b}},\un{\cl{p}})$ in phase space at $\cl{t}_0$ in both methods.
Hence, by normalizing the wave function with respect to the inner product \eqref{eq:inner_product_sqrt_kg_infty} we get:

\begin{equation}
\label{eq_bianchiI:adm_evolved_gaussian_wave_function}
\begin{aligned}
\Phi(t,\un{\beta}) =& \int_{\mathbb{R}^2} \frac{\dd^2 p}{ (2 \pi \hbar) \sqrt{2 \pi \sigma_+ \sigma_-}} \frac{e^{-\frac{i}{\hbar} \int_{\bar{\cl{t}}}^{\cl{t}} \dd \cl{t}' \sqrt{\mu^2(\cl{t}') + p_+^2 + p_-^2}}}{\sqrt[4]{\mu_1^2 + p_+^2 + p_-^2}} \\
&e^{\frac{i}{\hbar} \un{p} \cdot (\un{\beta}-\un{\cl{b}})} e^{- \frac{(\cl{p}_+ - p_+)^2}{2 \sigma^2_+}} e^{- \frac{(\cl{p}_- - p_-)^2}{2 \sigma^2_-}}.
\end{aligned}
\end{equation}
At this point we implement the BO approximation in \eqref{eq_bianchiI:adm_evolved_gaussian_wave_function}, which allow us to write 

\begin{equation}
\label{eq_bianchiI:rewriting_approx_adm_wave_packet}
\begin{aligned}
\Phi(\cl{t},\un{\beta}) \approx& \frac{e^{-\frac{i}{\hbar} \int_{\cl{t}_s}^{\cl{t}} \dd \cl{t}' \sqrt{\mu^2(\cl{t}')}}}{\sqrt[4]{\mu_1^2}} \int_{\mathbb{R}^2} \frac{\dd^2 p}{(2 \pi \hbar)\sqrt{2 \pi \sigma_+\sigma_-}} \\
&\frac{e^{- \frac{i}{2 \hbar} (p_+^2 + p_-^2) \int_{\cl{t}_s}^{\cl{t}} \dd t' \frac{1}{\sqrt{\mu^2(\cl{t}')}}}}{\sqrt[4]{1+ \frac{p_+^2 + p_-^2}{\mu_1^2}}} \\
&e^{\frac{i}{\hbar} \un{p} \cdot (\un{\beta} - \cl{\un{b}})} e^{- \frac{(\cl{p}_+ - p_+)^2}{2 \sigma^2_+}} e^{- \frac{(\cl{p}_- - p_-)^2}{2 \sigma^2_-}}.
\end{aligned}
\end{equation}
Comparing \eqref{eq_bianchiI:rewriting_approx_adm_wave_packet} with \eqref{eqn_bianchiI:final_wfv_fin} we see two main differences:
\begin{itemize}
\item the factor $\qty(1 + \frac{p_+^2 + p_-^2}{\mu_1^2})^{-\frac{1}{4}}$ in \eqref{eq_bianchiI:rewriting_approx_adm_wave_packet} is not present in \eqref{eqn_bianchiI:final_wfv_fin},  
\item the factor $(\mu^2_1)^{-\frac{1}{4}}$ in \eqref{eq_bianchiI:rewriting_approx_adm_wave_packet} is replaced by $(\mu^2(\cl{t}))^{-\frac{1}{4}}$ in \eqref{eqn_bianchiI:final_wfv_fin}. 
\end{itemize}

If we inspect the probability of finding the Universe in a region $\mathcal{B}$ of the plane $(\bp,\bim)$, we get the same result for both methods, namely:

\begin{equation}
\label{eq_bianchiI:probability_final}
\begin{aligned}
P(\mathcal{B}) =& \frac{\pi}{\sqrt{\frac{1}{4 \hbar^2} \qty(\int_{\cl{t}_s}^{\cl{t}} \dd \cl{t}' \frac{1}{\sqrt{\mu^2(\cl{t}')}})^2 + \frac{1}{4 \sigma^4_+}}} \\
&\frac{\pi}{\sqrt{\frac{1}{4 \hbar^2} \qty(\int_{\cl{t}_s}^{\cl{t}} \dd \cl{t}' \frac{1}{\sqrt{\mu^2(\cl{t}')}})^2 + \frac{1}{4 \sigma^4_-}}} \\
&\int_{\mathcal{B}} \frac{\dd^2 \beta}{(2\pi\hbar)^2 \pi \sigma_+ \sigma_-} e^{- \frac{\qty[\beta_+ - \qty(\cl{b}_++\cl{p}_+ \int_{\cl{t}_s}^{\cl{t}} \dd \cl{t}' \frac{1}{\sqrt{\mu^2(t')}})]^2}{\sigma_+^2 \qty[ \frac{1}{4 \hbar^2}\qty( \int_{\cl{t}_s}^{\cl{t}} \dd \cl{t}' \frac{1}{\sqrt{\mu^2(\cl{t}')}})^2 + \frac{1}{4 \sigma^4_+} ]}} \\
&e^{- \frac{\qty[\beta_- - \qty(\cl{b}_- +\cl{p}_- \int_{\cl{t}_s}^{\cl{t}} \dd \cl{t}' \frac{1}{\sqrt{\mu^2(t')}})]^2}{\sigma_-^2 \qty[ \frac{1}{4 \hbar^2}\qty( \int_{\cl{t}_s}^{\cl{t}} \dd \cl{t}' \frac{1}{\sqrt{\mu^2(\cl{t}')}})^2 + \frac{1}{4 \sigma^4_-} ]}}.
\end{aligned}
\end{equation}
The expectation value of the position operators $\hat{\un{\beta}}$ over the wave packet is 

\begin{equation}
\label{eq_bianchiI:mean_position}
\langle \hat{\un{\beta}} \rangle = \un{\cl{b}} + \un{\cl{p}} \int_{\cl{t}_s}^{\cl{t}} \dd \cl{t}' \frac{1}{\sqrt{\mu^2(\cl{t}')}}
\end{equation}
and the variances $\sigma^2_+(\cl{t})$ and $\sigma_-^2(\cl{t})$ change in time as follows

\begin{subequations}
\label{eq_bianchiI:width}
\begin{align}
&\sigma^2_\pm(\cl{t}) = \frac{\sigma_\pm}{\sqrt{2}} \sqrt{\frac{1}{4 \hbar^2}\qty( \int_{\cl{t}_s}^{\cl{t}} \dd \cl{t}' \frac{1}{\sqrt{\mu^2(\cl{t}')}})^2 + \frac{1}{4 \sigma^4_\pm}} ; \label{eq_bianchiI:width_+}\,.
\end{align}
\end{subequations}
The expression \eqref{eq_bianchiI:mean_position} coincide with the classical trajectories obtained by solving the equations \eqref{eq:motion_for_bp} and \eqref{eq:motion_for_bim} for

\begin{equation}
\label{eq_bianchiI:N_to_have_classical_motion}
N(\cl{t}) = \frac{3 c \kappa}{4 \pi G T} \frac{e^{3 \cl{t}}}{\sqrt{\mu^2(\cl{t})}}\,,
\end{equation}
while \eqref{eq_bianchiI:width_+} outlines that semiclassical wave packets do not spread in time as the Universe expands in the momenta representation. 

\section{Conclusions}
The aim of Quantum Cosmology is to provide a quantum description for homogeneous cosmological models. However, several problems arise. A major issue is the problem of time, {\it i.e.} the identification of a proper time-like variable avoiding the frozen formalism, which generically leads to inequivalent results whether it is realized before or after quantization (see for instance \cite{Guven1992}). In this work, we spell out this problem for a Bianchi I model. In particular, we made a comparison between the RPSQ formalism and the Vilenkin proposal. In the former the super-Hamiltonian constraint is classically solved, the isotropic variable is taken as time and only anisotropy degrees of freedom are quantized. In the latter, implementing both WKB and Born-Oppenheimer approximations, a probabilistic interpretation for the wave functions is achieved. The RPSQ is the most straightforward procedure, but it is plagued by the non-locality of the Hamiltonian density and the presence of a time-dependent mass-like term, due to the contribution of matter energy density. These issues can be solved for the Bianchi I model, suggesting a procedure which could be extended to more general cases. The Vilenkin proposal is more manageable, but it cannot be regarded as a fundamental approach, since a probabilistic interpretation is achieved only after performing a semiclassical limit. Hence, making the comparison with RPSQ we tested the viability of the Vilenkin proposal. We outlined how the two approaches provided the same probability densities for finding the Universe in a finite region of the $(\beta_+,\beta_-)$ plane if the spectra of the corresponding momenta extend over sufficiently small values, so that the contribution of anisotropies to the total energy density is negligible with respect to that of matter. In other words, the viability of Vilenkin proposal is restricted to those physical scenarios in which the overall dynamics is dominated by the matter contributions. However, it is worth noting that also the Dirac procedure for solving the constraint is, to some extent, a Born-Oppenheimer approximation on the system dynamics: indeed, selecting the Universe volume as the proper time variable makes it classical ab-initio and its role in the wave function is on a different footing with respect to the gravitational degrees of freedom $\beta_\pm$. The overlap of the probability distributions in the two cases above can then be regarded as the consequence of a similar Born-Oppenheimer approximation on the Dirac constraint, but it must be emphasized that only the reduced phase space approach is intrinsically valid for arbitrary values of the anisotropy variables. On this level, we can conclude that the main achievement of this analysis relies on the constraint we have to impose on the anisotropy variables phase space: the Vilenkin procedure holds only when a ``light dynamics'' of the anisotropies is considered.

{\em Acknowledgments.}---   
FC is supported by funds provided by the National Science Center under the agreement
DEC-2011/02/A/ST2/00294.

\end{document}